\documentclass[a4paper,twoside,titlepage,11pt]{article}
\usepackage{hyperref}

\usepackage{amssymb,amsmath,amsfonts} \usepackage{epsfig}

\newcommand{\be}{\begin{equation}} \newcommand{\ee}{\end{equation}}
\newcommand{\bea}{\begin{eqnarray}} \newcommand{\eea}{\end{eqnarray}}
\newcommand{\bino}[2]{\left( \begin{array}{c} #1 \\ #2 \end{array}
\right)}  
 \newcommand{\CN}{\mathcal{N}}

\newcommand{\id}{\hbox{1\kern-.27em l}}
\newcommand{\sid}{\hbox{\scriptsize1\kern-.27em l}}

\newcommand{\we}{\kern-.1em\wedge\kern-.1em}
\newcommand{\scal}{\kern-.13em\cdot\kern-.13em}

\newcommand{\II}{I\kern-.09em I}

\newcommand{\spa}{\ \ ,\ \ \ \ }

\newcommand{\Real}{\mathbb{R}}
\newcommand{\Pm}{\mathbb{P}}

\newcommand{\hti}{\tilde h}

\newcommand{\eq}{\begin{equation}}
\newcommand{\eqx}{\end{equation}}
\newcommand{\eqn}{\begin{eqnarray}}
\newcommand{\eqnx}{\end{eqnarray}}
\newcommand{\f}[2]{\frac{#1}{#2}}
\newcommand{\eps}{\varepsilon}
\newcommand{\tr}{\mbox{\rm Tr}\,}

\newcommand{\al}{\alpha}
\newcommand{\bt}{\beta}
\newcommand{\dl}{\delta}
\newcommand{\ff}{{\cal F}}
\newcommand{\nn}{{\cal N}}
\newcommand{\der}[2]{\f{\partial{#1}}{\partial{#2}}}
\newcommand{\gt}{\tilde{g}}

\newcommand{\ftl}{\tilde{f}}
\newcommand{\kk}{\hat{\cal K}}
\newcommand{\dhat}{\hat{\cal D}}
\newcommand{\rp}{\mathbb{R} \Pm^2}
\newcommand{\cor}[1]{\left\langle {#1} \right\rangle}

\begin{document}

\begin{titlepage}

\rightline{\vbox{ \hbox{\tt
 hep-th/0212069}}}
\vskip 1.3cm

\centerline{\LARGE \bf $SO(N)$ Superpotential, Seiberg-Witten Curves}
\centerline{ \LARGE \bf
and Loop Equations } \vskip 2cm
\centerline{{\Large   R. A. Janik${}^{a,b}$ and N. A. Obers${}^{a}$}
}
%\centerline{{\bf People}}
\vskip .8cm
\begin{center}
{\sl  ${}^a$ The
Niels Bohr Institute, Blegdamsvej 17, DK-2100 Copenhagen \O,
Denmark\\ ${}^b$ Jagellonian University, Reymonta 4, 30-059 Krakow, Poland  \\[.75ex] } \vskip .1in  {\small \sffamily janik@nbi.dk, obers@nbi.dk}
\end{center}
\vskip 0.8cm  \centerline{\bf \large Abstract} \vskip 0.1cm
\noindent

We consider the exact
superpotential of $\CN=1$ super Yang-Mills theory with gauge group
$SO(N)$ and arbitrary tree-level polynomial superpotential of one adjoint Higgs
field. A field-theoretic derivation of the glueball superpotential
is given, based on factorization of the $\CN=2$ Seiberg-Witten curve.
Following the conjecture of Dijkgraaf and Vafa, the result is matched
with the corresponding $SO(N)$ matrix model prediction.
The verification involves an explicit solution of the first
non-trivial loop equation, relating the spherical free energy to that
of the non-orientable surfaces with topology $\Real \Pm^2$.

\end{titlepage}

%-----------------------------------
%\tableofcontents  \setcounter{page}{1}

%\newpage
%--------------------------------------------

\section{Introduction}

Recently there has been a spectacular progress in the
understanding the low-energy dynamics of a large class of $\nn=1$
supersymmetric gauge theories. Although the proposal of Dijkgraaf
and Vafa (DV) linking effective superpotentials to random matrix
quantities arose from string theoretic reasonings
\cite{Dijkgraaf:2002fc,Dijkgraaf:2002vw,Dijkgraaf:2002dh},
its main part  has subsequently been proven by purely
field theoretic methods in \cite{Dijkgraaf:2002xd,Cachazo:2002ry}.

There has been much research in extending the DV framework to
accommodate matter in the fundamental representation
\cite{Argurio:2002xv,McGreevy:2002yg,Suzuki:2002gp,Bena:2002kw,Demasure:2002sc,Argurio:2002hk,Naculich:2002hr,Bena:2002ua},
exhibit Seiberg-Witten curves
\cite{Dorey:2002tj,Dorey:2002jc,Ferrari:2002jp,Fuji:2002wd,Berenstein:2002sn,Dijkgraaf:2002pp,Dorey:2002pq,Ferrari:2002kq,Naculich:2002hi,Feng:2002gb}, calculate
 gravitational couplings \cite{Klemm:2002pa,Dijkgraaf:2002yn}
 and various other related developments
  \cite{Chekhov:2002tf,Gorsky:2002uk,Aganagic:2002wv,Gopakumar:2002wx,
Tachikawa:2002wk,Tachikawa:2002ud}.

As pointed out in \cite{Cachazo:2001jy,Cachazo:2002pr} much of the physics of $\nn=2$
theories deformed by tree-level potentials can be effectively
obtained from the knowledge that the appropriate Seiberg-Witten (SW)
curve \cite{Seiberg:1994rs,Seiberg:1994aj}
for the undeformed theory factorizes. This approach has been
used in \cite{Ferrari:2002jp}
(see also \cite{Dijkgraaf:2002xd,Cachazo:2002ry}), together with the
Intriligator-Leigh-Seiberg (ILS)
linearity principle \cite{Intriligator:1994jr}
and `integrating-in' techniques to derive the random matrix DV
superpotential directly from properties of SW curves
for unitary groups.

The object of this paper is to i) calculate the effective glueball
superpotentials for orthogonal groups from the SW
perspective and then ii)  subsequently verify the DV conjecture
by deriving the same result from the random matrix
perspective using loop equation techniques. This is interesting as
these involve nonorientable graphs on the random matrix side and
especially as the original proposal of DV \cite{Dijkgraaf:2002dh} was somewhat
ambiguous (and indeed has to be slightly modified). Very recently,
as this work was in progress, there appeared papers  which
addressed the orthogonal groups from different perspectives,
namely perturbative methods \cite{Ita:2002kx} and CY/diagrammatic methods
\cite{Ashok:2002bi}.
The  interesting paper
\cite{Ashok:2002bi} has some overlap in that the loop equation is
also used there. We differ, however, on various aspects of the
derivation of the equation as well as the solution of it.

The plan of this paper is as follows. In section 2 we present the
factorization solution for SW curves for orthogonal
groups and use this information to  construct the
effective glueball superpotential by integrating-in $S$. Then we
discuss in section 3 how this  result should be reproduced from
a random matrix model following the DV proposal. This implies in
particular
a non-trivial relation between the spherical and $\Real \Pm^2$
contribution to the free energy, as was also noted in
\cite{Ita:2002kx,Ashok:2002bi}.
Finally, in section 4 we prove for arbitrary tree level potentials the
required random matrix identity using loop equation techniques.
We close the paper with a discussion.

%\newpage
\section{Exact $SO(N)$ superpotential from factorized SW curve}

In this section we give a field-theoretic derivation of
the exact superpotential in
$\CN=1$ super Yang-Mills theory with gauge group
$SO(N)$ and arbitrary polynomial superpotential of one adjoint Higgs
field.

We thus consider $\CN=2$ $SO(N)$ gauge theory broken to $\CN=1$
by a tree-level superpotential
\begin{equation}
\label{tree}
W_{\rm tree} = \sum_{p \geq 1} \frac{g_{2p}}{2p} {\rm Tr} \Phi^{2p}
%= \sum_{p \geq 1}  g_{p} v_p
\end{equation}
where the field $\Phi$ is in the adjoint. Note that, due to the
antisymmetry of $\Phi$ only even terms in the potential contribute.
Following the computation for $SU(N)$ in Ref~\cite{Ferrari:2002jp},
we compute here the exact superpotential for the $SO(N)$ case, in the
confining vacuum where  $\langle \Phi \rangle = 0$ classically.

To this end we use the ILS linearity principle \cite{Intriligator:1994jr}
which implies that under the addition of \eqref{tree}, the exact
effective superpotential is given by
\begin{equation}
\label{wq}
W_{\rm q} = \sqrt{2} \sum_{m=1}^r \tilde M_m M_m a^D_{m} (v_p,\Lambda)  + \sum_{p \geq 1}
g_{p} v_p
\end{equation}
Here, $\Lambda$ is the scale governing the running of the gauge coupling
constant, $M_{m}$, $\tilde M_m$ are the monopole fields,  $a^D_{m} $
are the dual $\CN=2$ $U(1)$ vector multiplet scalars,
$ r = [N/2] $ is the rank of $SO(N)$, and we have defined
\begin{equation}
\label{vp}
v_p \equiv \frac{1}{2p} \tr \Phi^{2p}
\end{equation}
Our aim is to obtain the universal superpotential $W (S,\Lambda)$ which
we achieve by first integrating out the monopole fields and subsequently
integrating in the $S$ field. Turning to the first step, the equation
of motion reads
\begin{equation}
\label{ad}
a^D_{m} (v_p,\Lambda) =0  \spa m = 1 \ldots r
\end{equation}
where  we assume that all species of monopoles condense.
The $a^D_{m}$ are given
by integrals of a meromorphic form over cycles of hyperelliptic curves
\cite{Seiberg:1994rs,Seiberg:1994aj}. In particular, for the case of
$SO(N)$ these were obtained in  \cite{Danielsson:1995is}
and \cite{Brandhuber:1995zp} for $N$ odd and even respectively,
and may be summarized according to
\begin{equation}
\label{curve}
y^2 = P(x)^2 - 4 x^{2q} \Lambda^{2 \tilde h} \spa
P(x) =  \prod_{k=1}^r (x^2 - e_k^2)
\end{equation}
Here $\tilde h = N-2$ is the dual Coxeter number of $SO(N)$ and
$q = 2r - \tilde h$ (hence $q=2$ for $SO(2N)$ and $q=1$ for $SO(2N+1)$).
The relation between the $v_p$ in \eqref{vp} and the moduli $e_k$ in
\eqref{curve} is
\begin{equation}
\label{vpa}
v_p = \frac{1}{p} \sum_{k=1}^r e_k^{2p}
\end{equation}
The vanishing of the $a^D_{m}$ implies a factorization constraint on
$P(x)$. For $SU(N)$ this was solved in \cite{Douglas:1995nw}, while in
the case of $SO(N)$ we find a similar solution%
\footnote{See also Refs.~\cite{Edelstein:2001mw,Fuji:2002vv,Feng:2002gb}.}
\begin{equation}
\label{pol}
P(x) = 2 x^q \Lambda^{\hti} T_{\hti} \left( \frac{x}{2\Lambda} \right)
\end{equation}
where $T_{l} (z) = \cos (l \arccos (z) ) $ is a Chebyshev polynomial.
Indeed, it is easy to check that for this choice the SW
curve factorizes
\begin{equation}
y^2 =  4^{q+1} \Lambda^{4r } z^{2q} (  z^2 -1)
[  U_{\hti -1} (z)]^2  \spa z \equiv \frac{x}{2\Lambda}
\end{equation}
For $SO(2N)$ this shows one six-fold, two single  and $2N-4$ double
zeroes, while for $SO(2N+1)$ there are two single and $2N-1$ double
zeroes.
Moreover, it is not difficult to check that the corresponding
meromorphic one-form is
\begin{equation}
\lambda = \frac{1}{2\pi i} (q P(x) - x P'(x) ) \frac{dx}{y}
   = -\frac{x^{q+1} \hti }{2\pi} \frac{dz}{\sqrt{1 -z^2}}
\end{equation}
exhibiting singularities at the single zeroes  $z = \pm 1$ of $y$ only.

{} From the solution \eqref{pol} we may now read off that at the
factorization point, the zeroes of $P(x)$ are $x=0$ and
\begin{equation}
e_k = 2 \Lambda \cos \left( \pi \frac{k-1/2}{\hti} \right) \spa k = 1 \ldots \hti
\end{equation}
Because of the symmetry $e_{\hti +1 -k} = - e_k$ this means that
the $r$ zeroes $e_k^2$ in \eqref{curve} are given by
\begin{equation}
e_k^2 = (2 \Lambda )^2 \cos^2 \left( \pi \frac{k-1/2}{\hti} \right) \spa
 k = 1, \ldots , r-1 \spa
e_r^2 = 0
\end{equation}
Substituting this in \eqref{vpa} then yields after some algebra
\begin{equation}
\label{vsol}
v_p (\Lambda^2) = \frac{\hti}{2p} \bino{2p}{p}  \Lambda^{2p}
\end{equation}
which thus specifies the point in the moduli space where the
monopoles coupling to each $U(1)$ in the Cartan subalgebra of
$SO(N)$ have become massless. It then follows from \eqref{wq} that
the exact superpotential is
\begin{equation}
\label{exsup}
W (\Lambda^2,g_{2p}) = \sum_{p \geq 1} g_{2p} v_p(\Lambda^2)
\end{equation}
We can now integrate in $S$ by Legendre transforming \eqref{exsup} using
\begin{equation}
\label{intin}
\frac{\partial W}{\partial \ln \Lambda^{N-2 }} = S
\end{equation}
which fixes the normalization\footnote{This choice was also adopted in
ref. \cite{VAFACS}.} of $S$.

The final result
for the exact $SO(N)$ superpotential is then
\begin{equation}
\label{nps}
 W_{\rm eff}(S,\Lambda^2,g_{2p}) = \frac{N-2}{2} \left[
 -S \ln ( \hat \Lambda(S)/\Lambda )^2
+  \sum_{p \geq 1} \frac{g_{2p}}{p} \bino{2p}{p} [\hat \Lambda(S)]^{2p}
\right]
\end{equation}
where the function $\hat \Lambda (S)$ is determined by the solution
of the equation
\begin{equation}
\label{Srel}
S = \sum_{p \geq 1} g_{2p}  \bino{2p}{p}\hat \Lambda^{2p}
\end{equation}
which follows from \eqref{intin}, using \eqref{exsup}, \eqref{vsol}.

At this point, it is useful to recall the corresponding result
for $SU(N)$ (or $U(N)$ with even potential) \cite{Ferrari:2002jp}
\begin{equation}
\label{nps2}
 W_{\rm eff}^{SU(N)}(S,\Lambda^2,f_{2p}) = N \left[
 -S \ln ( \hat \Lambda(S)/\Lambda )^2
+  \sum_{p \geq 1} \frac{f_{2p}}{2p} \bino{2p}{p} [\hat \Lambda(S)]^{2p}
\right]
\end{equation}
\begin{equation}
\label{Srelsu}
S = \frac{1}{2}\sum_{p \geq 1} f_{2p}  \bino{2p}{p}\hat \Lambda^{2p}
\end{equation}
where the tree-level potential is as in \eqref{tree} with coupling
constants $f_{2p}$.
We thus note the simple relation between the two cases
\begin{equation}
\label{sosurel}
W_{\rm eff}^{SO(N)} (g_{2p}) = \frac{N-2}{2N} W_{\rm eff}^{SU(N)}
(f_{2p} = 2 g_{2p})
\end{equation}
which will be relevant below.

\section{The Dijkgraaf-Vafa proposal}

Following the conjecture of Dijkgraaf and Vafa \cite{Dijkgraaf:2002dh}, we expect
to reproduce the exact $SO(N)$ superpotential \eqref{nps} from an
appropriate matrix model.
In this case we need to consider the partition function of
a one-matrix model with $\Phi$ in the adjoint representation of $SO(N)$,
i.e. real antisymmetric matrices\footnote{This is quite different
from the standard orthogonal ensemble in random matrix theory
where the matrices are real {\em symmetric}.}. We thus consider
\begin{equation}
\label{mm}
Z = \int d \Phi \exp  \left( - \frac{1}{g_s} \tr W_{\rm tree}(\Phi ) \right)
\spa S = g_s M
\end{equation}
where $W_{\rm tree} (\Phi)$ is the tree-level superpotential in \eqref{tree}
 and $M $  is the size of the matrices. Eliminating $g_s$, one may
 rewrite the partition function in the more standard random matrix
 model form
\eq
\label{e.rmz}
Z = \int \prod_{a>b} D\Phi_{ab} e^{-M \tr V(\Phi)}
\eqx
where $V(\Phi)=\sum_p \f{1}{2p} \gt_{2p} \Phi^{2p}$ and $\gt_{2p}$ is
related to the tree level potential coefficients through
$\gt_{2p}=g_{2p}/S$. The corresponding free energy in $Z=\exp( M^2 F)$
of this random matrix model has a $1/M$ expansion
\eq
\label{e.exp}
 F=\sum_{n=0}^\infty \f{1}{M^n} F_n
\eqx
For our purposes we will only be concerned with the contributions
$F_0$, $F_1$ arising from the sphere $S^2$ ($\chi=2$) and
the projective plane $\Real \Pm^2$ ($\chi =1$) respectively.
In terms of these quantities we extract the
free energies from \eqref{mm} according to
\begin{equation}
\label{Zexp}
\ff_{\chi =2} = - S^2 F_0 \spa
\ff_{\chi =1} = - S F_1
\end{equation}
where we have expanded $Z \simeq \exp ( - ( M^2/S^2) \ff_{\chi =2}
-(M/S) \ff_{\chi =1} )$.
According to the conjecture of \cite{Dijkgraaf:2002dh}
(in the form given in Refs.~\cite{Ita:2002kx,Ashok:2002bi} for $SO/Sp$),
the perturbative part of the superpotential is then given by
\begin{equation}
\label{conj}
W_{\rm pert}=  N \partial_S {\cal{F}}_{\chi =2} + 4 {\cal{F}}_{\chi
=1}
\end{equation}
in terms of the free energy contributions ${\cal{F}}_{\chi =1,2} $
defined in \eqref{Zexp}. Comparing with the exact result \eqref{nps}
obtained from factorization of the SW curve, we thus see that
in order for the conjecture to hold for all $N$ one needs
the relation
\begin{equation}
\label{frel}
\partial_S {\cal{F}}_{\chi =2} = -2 {\cal{F}}_{\chi =1}
%\partial_S t
\end{equation}
Moreover, given this relation, one should have that%
\footnote{All quantities $W$, ${\cal{F}}$, $F$ in this paper that do not carry
an explicit superscript referring to the group are for
$SO(N)$.}
\begin{equation}
\label{F2}
\partial_S {\cal{F}}_{\chi =2} (g) = \frac{1}{N-2 } W_{\rm pert} (g)
= \frac{1}{2N} W_{\rm pert}^{SU(N)} (f =2g) = \frac{1}{2}
\partial_S {\cal{F}}_{\chi =2}^{SU(N)} (f= 2 g)
\end{equation}
where we used the relation \eqref{sosurel} in the second step
and the last step follows from the DV conjecture for $SU(N)$.
Here, the arguments $g$ and $f=2g$ indicate the coupling constant
dependence.

In the next section we will use the loop equation to
prove the non-trivial identity \eqref{frel} relating the
$\Real \Pm^2$ free energy to the spherical contribution.
We will also derive the relation \eqref{F2}, after which
the proof of the $SO(N)$ conjecture \eqref{conj} immediately
follows from the one for $SU(N)$,
which was proven in \cite{Ferrari:2002jp,Dijkgraaf:2002xd,Cachazo:2002ry}.

\section{Loop equation}

In this section we will derive the result \eqref{frel} obtained by
factorization of the SW curve for the orthogonal
groups%
\footnote{The same result has  been recently obtained
using other methods \cite{Ita:2002kx,Ashok:2002bi}.  In Ref.~\cite{Ashok:2002bi}
the loop equation was used as well, but we differ considerably
on various aspects in the derivation.}
using the loop equation for the relevant
random matrix model. We also prove the relation \eqref{F2}
by comparing the zeroth order loop equation for $SO(N)$ and $SU(N)$.

The loop equation allows to find recursive relations among
the contributions in the $1/M$ expansion of the free
energy (see \eqref{e.exp}) of a random matrix model.
In practice the loop
equations do not involve the free energy directly but are rather
expressed in terms of its derivatives -- the resolvents
\eqn
\label{e.defres}
W(z) &\equiv& \cor{\f{1}{M}\tr \f{1}{z-\Phi}} =\f{1}{z} +\f{d}{dV(z)}
F\\
W(z,z) &\equiv& M^2 \left( \cor{\f{1}{M^2} \tr \f{1}{z-\Phi}\tr
\f{1}{z-\Phi} } - \cor{\f{1}{M}\tr \f{1}{z-\Phi}}^2 \right)
%=\f{d}{dV(z)} \f{d}{dV(z)} F
\eqnx
where $d/dV(z)$ is the loop insertion operator
\eq
\f{d}{dV(z)}=-\sum_{p=1}^\infty \f{2p}{z^{2p+1}} \der{}{\gt_{2p}}
\eqx
The $1/M$ expansion of the free energy (\ref{e.exp}) induces
the appropriate expansion of the resolvent $W(z)=W_0(z)+ (1/M)
W_1(z) + \ldots$. The form of the loop insertion operator then
determines the asymptotic large $z$ behavior of $W_i(z)$. In
particular we have $W_0(z) \sim 1/z$ and $W_i(z) \sim {\cal
O}(1/z^2)$ for $i>0$. These conditions in general ensure the
uniqueness of the solution of the loop equations.

The loop equation is derived by requiring the invariance of the
partition function
(\ref{e.rmz}) under the coordinate reparameterization
\eq
\label{e.coord}
\Phi=\Phi'- \eps \sum_{k=0}^\infty \f{\Phi'^{2k+1}}{z^{2k+2}}
\eqx
The transformation properties of
the measure follow from
\eq d\Phi_{ab}=d\Phi_{ab}' -\eps
\sum_{k=0}^\infty \sum_{l=0}^{2k} \f{\Phi'^l_{ac} d\Phi'_{cd}
\Phi'^{2k-l}_{db}}{z^{2k+2}}
\eqx
This is the initial starting
point of ref. \cite{Ashok:2002bi} but from now on our treatment differs
considerably.

The Jacobian matrix for the reparameterization \eqref{e.coord} is
\eq
\f{\partial \Phi_{ab}}{\partial
\Phi'_{ij}} =\dl_{ai}\dl_{bj} -\eps \sum_{k=0}^\infty
\sum_{l=0}^{2k} \f{1}{z^{2k+2}} \left[ \Phi'^l_{ai}
\Phi'^{2k-l}_{jb} -\Phi'^l_{aj} \Phi'^{2k-l}_{ib} \right]
\eqx
where $a>b$ and $i>j$. The resulting Jacobian is then obtained from
\eq
J=1-\eps \sum_{a>b}
\sum_{k=0}^\infty \sum_{l=0}^{2k} \f{1}{z^{2k+2}} \left[
\Phi'^l_{aa} \Phi'^{2k-l}_{bb} -\Phi'^l_{ab} \Phi'^{2k-l}_{ab}
\right]
\eqx
Using the symmetry with respect to the interchange of
$a$ and $b$ we have $\sum_{a>b}=\f{1}{2}\sum_{a,b}$. Furthermore
since $\Phi'^l_{ab}=(-1)^l \Phi'^l_{ba}$ we can rewrite $J$ in a
compact form
\eq
J=1-\f{\eps}{2} \left( \tr \f{1}{z-\Phi'}
\right)^2+ \f{\eps}{2} \tr \f{1}{z^2-\Phi'^2} \equiv 1-\f{\eps}{2}
\left( \tr \f{1}{z-\Phi'} \right)^2+ \f{\eps}{2} \f{1}{z} \tr
\f{1}{z-\Phi'}
\eqx
Combining this with the transformation
property of the potential $\tr V(\Phi) =\tr V(\Phi') -\eps \tr
\f{V'(\Phi')}{z-\Phi'}$ the loop equation follows from the
invariance property
\eq
\int D\Phi e^{-M\tr V(\Phi)} \equiv \int
D\Phi' J\cdot  \left(1+M \eps \tr \f{V'(\Phi')}{z-\Phi'}\right)
e^{-M\tr V(\Phi')} =\int D\Phi' e^{-M \tr V(\Phi')}
\eqx
Standard manipulations then yield
\eq
\label{loopeq}
\f{1}{2} \left(W(z)^2 +\f{1}{M^2}
W(z,z) \right) -\f{1}{M} \f{1}{2z} W(z) -\int_C \f{dw}{2\pi i}
\f{V'(w)}{z-w} W(w)=0
\eqx
Let us denote by $\kk$ the integral operator
\eq
\label{e.defkk}
\kk f(z) =\int_C \f{dw}{2\pi i} \f{V'(w)}{z-w} f(w)
\eqx
Then it follows from \eqref{loopeq} that the equations for the
$S^2$ and $\rp$ contributions to the resolvent are respectively
\eqn
\label{e.genz}
\kk W_0(z) &=&\f{1}{2} W_0^2(z) \\
\label{e.rp} \kk W_1(z) &=&W_0(z) \left(W_1(z) -\f{1}{2z} \right)
\eqnx
These equations should be solved subject to the asymptotic
conditions $W_0(z) \sim 1/z$ for $z\to \infty$ and $W_1(z) \sim
{\cal O}(\f{1}{z^2})$.

\subsubsection*{$S^2$ topology}

It is convenient to relate the solution of the genus 0 loop equation
(\ref{e.genz}) to the result for matrix models relevant for the
unitary case. In that case, the relevant equation is
\cite{Ambjorn:1993gw,Ambjorn:1990ji}
\eq
\kk^{SU} W_0^{SU}(z) = (W_0^{SU}(z))^2
\eqx
where $\kk^{SU}$ is given by the same formula as (\ref{e.defkk}) but
with the coupling constants $\gt_{2p}$ substituted by coupling
constants of the complex matrix model $\ftl_{2p}$.
Comparing the two we see that the orthogonal resolvent $W_0 \equiv
W_0^{SO}$ is equal to the `unitary' resolvent calculated with the
couplings $\ftl_{2p}=2 \gt_{2p}$. Using the defining relation
(\ref{e.defres}) we see that this implies that
\eq
\label{sosurel2}
F_0^{SO}(\gt)=\f{1}{2} F_0^{SU}(\ftl=2\gt)
\eqx

\subsubsection*{$\rp$ topology}

In order to find the solution for $W_1(z)$ it is convenient to
introduce the function $\dhat W_0(z)$ where $\dhat$ is the
differential operator
\eq
\label{Dhat}
\dhat= \sum_p \gt_{2p}
\f{\partial}{ \partial \gt_{2p}  } =
\sum_p g_{2p}
\f{\partial}{\partial g_{2p}}
\eqx
\eq
\label{comrel}
[\dhat , \kk ] = \kk \spa [ \dhat, \frac{d}{dV} ] = - \frac{d}{dV}
\eqx
Note that this differential
operator does not depend on whether the couplings $\{g_{2p}\}$ are
rescaled by $S$ or not. Acting with $\dhat$ on the genus 0
equation (\ref{e.genz}) one obtains
\eq
\kk (\dhat W_0)= W_0 \dhat
W_0 -\f{1}{2} W_0^2
\eqx
where we used the first commutator in \eqref{comrel}.
In terms of these quantities we can find
a solution of (\ref{e.rp}) in the form
\eq
W_1(z)=\al \left(
W_0-\f{1}{z} \right) +\bt \dhat W_0
\eqx
where we recall the condition $W_1(z) \sim {\cal O}(\f{1}{z^2})$.
Substituting this back
into (\ref{e.rp}) yields uniquely $\al=-\bt=-\f{1}{2}$ thus
\eq
W_1=-\f{1}{2} \left(W_0 -\f{1}{z} -\dhat W_0\right) \equiv
-\f{1}{2} \f{d}{dV(z)} \left(2F_0 -\sum_p \gt_{2p}
\f{\partial}{\partial \gt_{2p}} F_0 \right)
\eqx
where we used that $\kk \frac{1}{z} = 0$ and
the second commutator in \eqref{comrel}.
Since $W_1=dF_1/dV(z)$ we can uniquely reconstruct $F_1$ from the above equation
up to an inessential coupling constant independent
additive constant. This yields finally
\eq
\label{e.fone}
F_1=-\f{1}{2} \left(2 -\sum_p \gt_{2p} \f{\partial}{\partial \gt_{2p}}
\right) F_0
\eqx

\subsubsection*{Link with the DV proposal}

Let us now reinterpret the matrix model identities obtained in the
previous section within the gauge theoretical framework.

We first examine the perturbative expansion of $F_0$:
\eq
F_0=\sum_{\{n_{2p}\}} a_{\{n_{2p}\}} \prod_p
\gt_{2p}^{n_{2p}}= \sum_{\{n_{2p}\}} a_{\{n_{2p}\}} \prod_p
g_{2p}^{n_{2p}} S^{-\sum n_{2p}}
\eqx
where the sum runs over the
various possible numbers of vertices of different types and
$a_{\{n_{2p}\}}$ are the relevant combinatorial factors.
A crucial property is now that the differential operator $\partial_S$
can be directly related to the operator $\dhat$ in the matrix model.
In particular, using the
fact that $ \ff_{\chi=2}= -S^2 F_0$ we  easily get
\eq
\der{\ff_{\chi=2}(S)}{S} = -\sum_{\{n_{2p}\}} \left(2- \sum n_{2p}\right)
a_{\{n_{2p}\}} \prod_p g_{2p}^{n_{2p}} S^{1-\sum n_{2p}}
\eqx
The above can be rewritten in terms of random matrix quantities as
\eq
\der{\ff_{\chi=2}(S)}{S}= -S \left(2 -\sum_p g_{2p} \der{}{g_{2p}}
\right) F_0
\eqx
Using \eqref{Dhat} and the fact that $\ff_{\chi=1}(S)=-S F_1$, the
solution of the loop equation (\ref{e.fone}) thus implies
\eq
\der{\ff_{\chi=2}(S)}{S}= -2 \ff_{\chi=1} (S)
\eqx
which verifies the announced identity \eqref{frel}.
Turning to the other relation \eqref{F2},
we note that the matrix model couplings for the unitary and orthogonal
groups were related to the gauge theoretical couplings
via $\ftl_{2p}=2\gt_{2p}$ which before the rescaling is equivalent to
 $f_{2p}= 2 g_{2p}$.
In particular it then follows from \eqref{sosurel2} that
$\ff_{\chi=2}^{SO} (g) =\f{1}{2} \ff_{\chi=2}^{SU} (f = 2g)$
which proves \eqref{F2}.

The effective potential obtained in section 2 from the Seiberg-Witten
curve for orthogonal groups was shown to be equal to
\eq
W_{\rm eff}=\f{N-2}{2} \der{\ff_{\chi=2}^{SU} (f=2g)}{S}
\eqx
This can be rewritten in terms of the quantities related to the
orthogonal matrix model as
\eq
W_{\rm eff}=(N-2) \der{\ff_{\chi=2}^{SO}}{S}=N \der{\ff_{\chi=2}^{SO}}{S}
+ 4 \ff_{\chi=1}^{SO}
\eqx
The multiplicative factor 4 has been first identified in
\cite{Ita:2002kx} where it was shown to arise from the fact that the
field theoretical determinants gave 1 for graphs with topology of
$S^2$ and 4 for graphs with the topology of $\rp$.

\section{Conclusions}

We have performed a field-theoretic computation of the exact
superpotential for $\CN=1$ $SO(N)$ gauge theory with arbitrary
tree-level potential of an adjoint field. Here we used the factorization
properties of SW curves for the orthogonal groups together with
the ILS linearity principle. Comparison of this
result with the matrix model conjecture \cite{Dijkgraaf:2002dh} (see also
\cite{Ita:2002kx,Ashok:2002bi}) implied the existence of a non-trivial identity
relating the spherical free energy to that of the next
contribution on $\rp$. By explicitly solving the loop
equation, we have been able to derive this identity for
arbitrary potential. Moreover, the zeroth order loop equation
enabled us to relate the spherical part in the $SO(N)$ theory
to that of the $SU(N)$ theory, thereby showing that
the validity of the $SU(N)$ conjecture directly implies the
corresponding one for $SO(N)$.

The fact that certain
quantities in supersymmetric gauge theories apparently know about
information encoded in the loop equation, may be regarded as
further evidence for the deep connection between these
gauge theories and matrix models. It is also interesting to note
that, in the end, the $SO(N)$ superpotential can be expressed
in terms of the corresponding $SU(N)$ planar free energy.
Our work also lends further support for the conjecture
\cite{Ferrari:2002jp}
that the ILS hypothesis is intimately related to the DV matrix
theory proposal.

One obvious generalization is to apply the analysis of this note
to the symplectic case. In particular, it would be interesting
to use the methods of section 4 to derive from the loop equation the
expected relation $\partial_S \ff_{\chi =2} = 2 \ff_{\chi =1}$.
Another area that has so far not received much attention is to
consider the addition of matter in the fundamental representation to
these $\CN=1$ $SO(N)$ gauge theories. Here, the appearance of
different types of $\chi=1$ contributions might give interesting
results, while a possible connection to the case of $SU(N)$ with
fundamental matter might generate further insights as well.

\bigskip

\noindent{\bf Acknowledgments} We would like to thank Robbert Dijkgraaf
for helpful discussions.\\ RJ was supported by the EU network on
``Discrete Random Geometry'' and KBN grant 2P03B09622.

\addcontentsline{toc}{section}{References}

%The following two lines is for bibtex only:
%\bibliographystyle{utphys}
%\bibliography{bibliodv}
%\bibliographystyle{../INPUT/utphys}
%\bibliography{../BIB/bibliodv}

%\end{document}
\providecommand{\href}[2]{#2}\begingroup\raggedright\endgroup

\end{document}